\documentclass[secnumarabic,amssymb,nobibnotes,aps,prd]{revtex4-2}
\usepackage[latin9]{inputenc}
\usepackage{bm}
\usepackage{amsmath}
\usepackage{graphicx}
\usepackage{esint}



\begin{document}
\title{Energetic particle modes in burning plasmas}
\author{Yang Li}
\email{leeyang_swip@outlook.com}

\address{Southwestern Institute of Physics, PO Box 432, Chengdu 610041,
People\textquoteright s Republic of China}
\begin{abstract}
In this work, we present a novel kind of energetic particle modes
with frequencies in the range of thermal ion diamagnetic and/or transit
frequencies in burning plasmas. It is shown that the continuum structure
can be broken by the energetic particles. In addition, the mode destabilization
does not require the energetic particle drive to overcome the continuum
damping, which indicates that they are much easier to be excited and
thus manifest themselves as wide-band modes. Based on the analysis,
a switch or even an oscillation between phases of discrete eigenmodes
and wide-band modes can be predicted.
\end{abstract}
\maketitle
Energetic particles (EPs), as the product of fusion reaction and auxiliary
heating, play a key role in magnetically confined fusion devices.
Such EPs are thermalized by Coulomb collision, which provides the
energy source for a sustainable igniting fusion reactor. However,
EPs can also resonantly excite collective modes such as shear Alfvén
waves (SAWs) which lead to anormalous EP loss\citep{chen_theory_2007}.
The EP induced SAWs have been investigated extensively\citep{chen_physics_2016}.
In previous studies, the density of EPs is considered to be low (about
$O\left(\beta^{3/2}\right)$ of the thermal ion density), especially
in the analysis of inertial layer\citep{zonca_kinetic_1996}, where
$\beta$ is the ratio of plasma energy to magnetic field energy. Such
condition is usually valid when only auxiliary heating is taken into
account. In future tokamaks such as ITER or even fusion reactors,
however, the product of D-T reaction together with power injection
will probably lead to a higher proportion of EPs\citep{aymar_iter_2002}. Thus, it is necessary
to consider the effect of higher EP proportion on the SAW excitation.
In this letter, a novel mechanism for the excitation of SAWs by high
proportion of EPs in burning plasmas is presented. 

The SAWs usually exhibit themselves as discrete eigenmodes, such as
Beta-induced Alfvén Eigenmode (BAE) \citep{heidbrink_observation_1993,turnbull-global-1993}etc.,
due to magnetic field geometry and non-uniformities. These discrete
modes can only exist in the gaps of SAW continuum spectrum since the
modes within the continuum are suppressed by the continuum damping
mechanism\citep{chen_plasma_1974}. In theoretical study, the continuum
spectrum is determined by the property of the so-called inertia, which
is calculated from the asymptotic behavior of the solution of the
vorticity equation in inertial layer. If the resonant drive from EPs
exceeds the threshold of the damping of continuum spectrum, energetic
particle modes (EPMs) are then excited\citep{chen_theory_1994}.

For low frequency SAWs, when the EP effect is neglected in the inertial
layer due to low density, the vorticity equation in inertial layer
is then periodic. Here low frequency modes are at the frequency range
of beta-induced Alfvén eigenmode (BAE) and kinetic balooning mode
(KBM)\citep{Tang_kinetic_1980,cheng_kinetic_1982}. However, if the
density of EPs increases from $O\left(\beta^{3/2}\right)$ to $O\left(\beta\right)$
of the thermal ions, the EP effect should be taken into account in
the inertial layer analysis. In this letter, it will be demonstrated
that the EP effect, in inertial layer, breaks the periodicity of the
vorticity equation and the structure of continuum spectrum, which
makes the EPMs much more unstable. It is found that there will not
be any threshold for EP drive to excite EPMs. 

In this work, the standard gyrokinetic model is applied and the perturbed
distribution function can be expressed as\citep{chen_kinetic_1991,antonsen_kinetic_1980}
\begin{align}
\delta f_{s} & =\left(\frac{e}{m}\right)_{s}\left[\frac{\partial F_{0}}{\partial{\cal E}}\delta\phi-\frac{QF_{0}}{\omega}e^{iL_{k}}J_{0}\left(k_{\perp}\rho_{s}\right)\delta\psi\right]_{s}\nonumber \\
 & +\left(\frac{e}{m}\right)_{s}\frac{\partial F_{0s}}{B\partial\mu}\left[\left(1-J_{0}\left(k_{\perp}\rho_{s}\right)e^{iL_{k}}\right)\left(\delta\phi+i\frac{v_{\parallel}}{\omega}\mathbf{b}\cdot\nabla\psi\right)-\frac{v_{\perp}}{k_{\perp}c}J_{1}\left(k_{\perp}\rho_{s}\right)e^{iL_{k}}\delta B_{\parallel}\right]_{s}\nonumber \\
 & +e^{iL_{ks}}\delta K_{s},\label{eq:gyro_f}
\end{align}
where $\delta K_{s}$ satisfies
\begin{align}
\left[\omega_{tr}\partial_{\vartheta}-i\left(\omega-\omega_{d}\right)\right]_{s}\delta K_{s} & =i\left(\frac{e}{m}\right)_{s}QF_{0s}\left[J_{0}\left(k_{\perp}\rho_{s}\right)\left(\delta\phi-\delta\psi\right)\right]\label{eq:gyro-k}\\
+ & i\left(\frac{e}{m}\right)_{s}QF_{0s}\left[\left(\frac{\omega_{d}}{\omega}\right)_{s}J_{0}\left(k_{\perp}\rho_{s}\right)\delta\psi+\frac{v_{\perp}}{k_{\perp}c}J_{1}\left(k_{\perp}\rho_{s}\right)\delta B_{\parallel}\right],\nonumber 
\end{align}
where $\delta\phi$ and $\delta B_{\parallel}$ are, respectively,
the perturbed electrostatic potential and the parallel magnetic field,
$\delta\psi$ is related to the parallel perturbed vector potential
$\delta A_{\parallel}$ by $\delta A_{\parallel}=-i\left(c/\omega\right)\mathbf{b}\cdot\nabla\delta\psi$,
$\omega$ is mode frequency, $\bm{b}=\bm{B}/B$, the subscript $s$
indicates particle species except for core electrons which are assumed
to be adiabatic; i.e., $s=E,i$ for EPs and thermal ions respectively,
$e_{s}$ and $m_{s}$ are the electric charge and mass for the species,
$F_{0s}$ is the equilibrium distribution function, ${\cal E}=v^{2}/2$
the energy per unit mass, $\mu=v_{\perp}^{2}/\left(2B\right)$ the
magnetic moment, $QF_{0s}=\left(\omega\partial_{{\cal E}}+\hat{\omega}_{*}\right)_{s}F_{0s}$,
$\hat{\omega}_{*s}F_{0s}=\omega_{cs}^{-1}\left(\mathbf{k}\times\mathbf{b}\right)\cdot\nabla F_{0s}$,
$\omega_{cs}=e_{s}B/\left(m_{s}c\right)$ is the cyclotron frequency,
$\mathbf{k}=-i\nabla$ is the wave vector, $J_{0}$ and $J_{1}$ are
the 0th and 1st order Bessel functions of the first kind, $k_{\perp}$
the perpendicular to $\mathbf{b}$ wave vector, $\rho_{Ls}=m_{s}cv_{\perp}/e_{s}B$
the Larmor radius, $L_{ks}=\omega_{cs}^{-1}\left(\mathbf{k}\times\mathbf{b}\right)\cdot\mathbf{v}$
is the generator of coordinate transformation from the guiding center
to particle variables, $\omega_{tr}=v_{\parallel}/qR_{0}$ the transit
frequency with $q$ being the safety factor and $R_{0}$ being the
major radius, $\omega_{ds}=\omega_{cs}^{-1}\left(\mathbf{k}\times\mathbf{b}\right)\cdot\left(\mu\nabla B+v_{\parallel}^{2}\mathbf{\kappa}\right)$
is the magnetic drift frequency and $\kappa=\mathbf{b}\cdot\nabla\mathbf{b}$
is the magnetic field curvature. Here the ballooning representation
\citep{connor_shear_1978} is already considered and $\vartheta$
is the extended poloidal angle. With the $s-\alpha$ model\citep{connor_shear_1978},
it can be expressed that $\omega_{ds}=g\left(\vartheta\right)k_{\vartheta}c\left(v_{\perp}^{2}/2+v_{\parallel}^{2}\right)/q_{s}BR_{0}$,
$g\left(\vartheta\right)=\cos\vartheta+\left(s\vartheta-\alpha\sin\vartheta\right)\sin\vartheta$
and $\alpha=-R_{0}q^{2}\beta^{\prime}$with a prime denoting derivation
with respect to $r$. 

Finite $\beta$ tokamak plasmas are considered and $\varepsilon\sim\beta^{1/2}$
is used as small parameter for formal orderings. And $\varepsilon\sim L_{p}/R_{0}$,
where $L_{p}$ is the radial inhomogeneity scale length. A higher
density of EPs is considered as $n_{E}\sim\varepsilon^{2}n_{i}$ with
temperature $T_{E}\sim\varepsilon^{-2}T_{i}$. This is suitable for future fusion reactor with 3.5Mev $\alpha$ particles and some KeV thermal ions\citep{aymar_iter_2002}. The frequency orderings
are given as $\omega_{*pi}\varepsilon^{-1}\sim\omega_{ti}\varepsilon^{-1}\sim\omega\varepsilon^{-1}\sim\omega_{A}\sim\omega_{tE}\sim\varepsilon\omega_{*pE}$,
where $\omega_{*ps}=\omega_{*ns}+\omega_{*Ts}$ , $\omega_{*ns}=\left(cT_{s}/e_{s}B\right)\left(\mathbf{k}\times\mathbf{b}\right)\cdot\nabla n_{s}/n_{s}$,
$\omega_{*Ts}=\left(c/e_{s}B\right)\left(\mathbf{k}\times\mathbf{b}\right)\cdot\nabla T_{s}$,
$\omega_{ts}=\sqrt{2T_{s}/m_{s}}/qR_{0}$, $\omega_{A}=v_{A}/qR_{0}$
is the Alfvén frequency and $v_{A}=B/\sqrt{4\pi n_{i}m_{i}}$ is the
Alfvén velocity. It should be noted that the desity of EPs is still
much smaller than that of the thermal ions and the Alfvén velocity
is determined by the thermal ions. The wavelength orderings can be
written as $k_{\vartheta}\rho_{E}\sim\varepsilon$ and $k_{\vartheta}\rho_{i}\sim\varepsilon^{2}$.
And $\omega_{ds}\sim qk_{\perp}\rho_{s}\omega_{ts}$, noting again
$s=E,i$. 

In order to study the SAW dispersion relation, the problem can be
determined by a close set of governing equations including the quasineutrality
equation, the perpendicular Ampère's law and the vorticity equation.
By assuming that the electron response is adiabatic, the quasineutrality
equation can be given as
\begin{equation}
\left(\frac{e_{i}n_{i}}{T_{i}}+\frac{en_{i}}{T_{e}}\right)\left(\delta\phi-\delta\psi\right)+\frac{en_{i}}{T_{i}}\left(1-\frac{\omega_{*pi}}{\omega}\right)b_{i}\delta\psi=\left<\sum_{s=i,e}J_{0}\delta K_{s}\right>,\label{eq:quasin}
\end{equation}
where $\left\langle \cdots\right\rangle =2\pi\sum_{v_{\parallel}/\left|v_{\parallel}\right|}\int\left(\cdots\right)Bd\mu d{\cal E}/\left|v_{\parallel}\right|$,
$b_{s}=k_{\perp}^{2}c^{2}m_{s}T_{s}/e^{2}B^{2}$ and the density of
electron is approximately equal to the thermal ion density since $n_{E}/n_{i}\sim\varepsilon^{2}$.
The perpendicular Ampère's law can be cast as
\begin{equation}
\delta B_{\parallel}=\frac{4c\pi\delta\psi}{\omega B^{2}}\left(\bm{k}\times\bm{b}\right)\cdot\nabla P-\sum_{s}4\pi m_{s}\left\langle \frac{2J_{1}}{k_{\perp}\rho_{s}}\mu\delta K_{s}+\frac{e_{E}}{m_{E}}\frac{QF_{0s}}{\omega}\left(1-\frac{2J_{1}J_{0}}{k_{\perp}\rho_{s}}\right)\mu\delta\psi\right\rangle .\label{eq:perp-AL}
\end{equation}
Here argument of $J_{0}$ and $J_{1}$ is $k_{\perp}\rho_{s}$. The
equilibrium distribution function of EPs is regarded as isotropic
since the EPs are mainly generated by fusion reaction. $P$ is the
total pressure including both thermal particles and EPs. By integrating
Eq.(\ref{eq:gyro-k}) with multiper $4\pi ei\omega J_{0}/k_{\vartheta}^{2}c^{2}$and
following the procedures in Ref.\citep{chen_kinetic_1991}, the vorticity
equation can be obtained as
\begin{align}
B\bm{b}\cdot\nabla\frac{k_{\perp}^{2}}{Bk_{\vartheta}^{2}}\bm{b}\cdot\nabla\delta\psi+\frac{\omega^{2}}{v_{A}^{2}}\left(1-\frac{\omega_{*pi}}{\omega}\right)\frac{k_{\perp}^{2}}{k_{\vartheta}^{2}}\delta\phi+\frac{\alpha g\left(\vartheta\right)}{q^{2}R_{0}^{2}}\delta\psi+\frac{\omega\alpha}{2ck_{\vartheta}q^{2}R_{0}}\delta B_{\parallel} & =\left<\sum_{s=i,E}\frac{4\pi e_{s}}{k_{\vartheta}^{2}c^{2}}J_{0}\omega\omega_{ds}\delta K_{s}\right>\nonumber \\
+\frac{4\pi\omega^{2}}{k_{\vartheta}^{2}c^{2}}\left\langle e_{E}\mu\frac{QF_{0E}}{\omega}\left(1-\frac{2J_{1}J_{0}}{k_{\perp}\rho_{E}}\right)\delta B_{\parallel}\right\rangle +\left\langle \frac{4\pi}{k_{\vartheta}^{2}c^{2}}\left(1-J_{0}^{2}\right)\omega\omega_{dE}\frac{e_{E}^{2}}{m_{E}}\frac{QF_{0E}}{\omega}\right\rangle \delta\psi & +\frac{4\pi\omega^{2}}{k_{\vartheta}^{2}c^{2}}\left\langle \frac{e_{E}^{2}}{m_{E}}\frac{QF_{0E}}{\omega}\left(1-J_{0}^{2}\right)\right\rangle \delta\phi.\label{eq:vorticity}
\end{align}
Note that here the parallel Ampère's law is applied,
\begin{equation}
\frac{k_{\perp}^{2}}{k_{\vartheta}^{2}}\bm{b}\cdot\nabla\delta\psi=\frac{4\pi}{k_{\vartheta}^{2}c^{2}}i\omega\left\langle \sum_{s}e_{s}v_{\parallel}\delta f_{s}\right\rangle .\label{eq:para-AL}
\end{equation}

To gain further insights of Eqs. (\ref{eq:quasin})-(\ref{eq:vorticity}),
the two-scale approach can be applied. The problem is seperated into
the large $\left|\vartheta\right|$ region (or inertial layer) and
the moderate $\left|\vartheta\right|$ region (or ideal region). In
the inertial layer, the mode structure varies on the large scale $\left|\vartheta_{1}\right|\sim\varepsilon^{-1}$
and the small scale $\left|\vartheta_{0}\right|\sim1$. The equation
set above can be solved order by order. In inertial layer, Larmor
radius parameters order as $k_{\perp}\rho_{i}\sim\varepsilon$ and
$k_{\perp}\rho_{E}\sim1$. By redefining the perturbed field as $\delta\Phi=k_{\perp}\delta\phi/k_{\vartheta}$,
$\delta\Psi=k_{\perp}\delta\psi/k_{\vartheta}$ and $\delta\hat{B}_{\parallel}=k_{\perp}\delta B_{\parallel}/k_{\vartheta}$,
the solutions of perturbed fields and distribution function can be
described for instance,
\begin{equation}
\delta\Psi=\delta\Psi^{\left(0\right)}+\varepsilon\delta\Psi^{\left(1\right)}+\varepsilon^{2}\delta\Psi^{\left(2\right)}+\cdots,
\end{equation}
\begin{equation}
\delta K_{s}=\delta K_{s}^{\left(0\right)}+\varepsilon\delta K_{s}^{\left(1\right)}+\varepsilon^{2}\delta K_{s}^{\left(2\right)}+\cdots.
\end{equation}
The 0-th order solutions of the governing equations can be obtained
as $\delta K_{i}^{\left(0\right)}=0$, $\delta\Phi^{\left(0\right)}=\delta\Psi^{\left(0\right)}$,
$\partial_{\vartheta_{0}}^{2}\delta\Psi^{\left(0\right)}=0,$ 
\begin{equation}
\delta K_{E}^{\left(0\right)}\simeq\frac{e_{E}}{\omega m_{E}}QF_{0E}\frac{k_{\vartheta}}{k_{\perp}}J_{0}\delta\Psi^{\left(0\right)},\label{eq:kE0}
\end{equation}
\begin{equation}
\delta\hat{B}_{\parallel}^{\left(0\right)}\simeq\frac{ck_{\vartheta}}{\omega q^{2}R_{0}}\frac{\alpha_{c}}{2}\delta\Psi^{\left(0\right)},\label{eq:Bpara0}
\end{equation}
where $\alpha=\alpha_{E}+\alpha_{c}$ includes both thermal particles
($\alpha_{c}$) and EPs components ($\alpha_{E}$). Note that $\delta\Psi^{\left(0\right)}$
is independent of $\vartheta_{0}$. The solution of $\delta B_{\parallel}^{(1)}$
will not be necessary in later analysis and the related 1st order
perpendicular Ampère's law is not presented here. By decomposing $\left[\delta\Psi^{(1)},\delta\Phi^{(1)}\right]=\left[\delta\Psi_{s},\delta\Phi_{s}\right]\sin\vartheta_{0}+\left[\delta\Psi_{c},\delta\Phi_{c}\right]\cos\vartheta_{0}$
and matching the Fourier component in $\vartheta_{0}$, the solution
of 1st order of the governing equations can be obtained as\citep{zonca_kinetic_1996}
$\delta\Psi_{s}\simeq-s\vartheta_{1}\alpha_{E}\delta\Psi^{\left(0\right)}/2\left|s\vartheta_{1}\right|^{2}$,
$\delta\Psi_{c}\simeq\delta\Psi_{s}/\left(s\vartheta_{1}\right)$,
$\delta\Phi_{s}=s\vartheta_{1}\delta\Phi_{c}$, 
\begin{equation}
\frac{\delta\Phi_{c}-\delta\Psi_{c}}{\delta\Psi^{\left(0\right)}}=-\frac{2cT_{i}}{eB_{0}}\frac{k_{\vartheta}}{\omega R_{0}}\frac{N\left(qR_{0}\omega/v_{ti}\right)}{D\left(qR_{0}\omega/v_{ti}\right)},\label{eq:phi-1}
\end{equation}
\begin{align}
\delta K_{i}^{\left(1\right)} & =i\frac{k_{\vartheta}}{k_{\perp}}\frac{\left(e/m\right)_{i}QF_{0i}}{\omega^{2}-\omega_{ti}^{2}}\left[\left(i\omega+s\vartheta_{1}\omega_{ti}\right)\frac{k_{\vartheta}\Omega_{di}}{\omega}\delta\Psi^{\left(0\right)}+i\omega\left(\delta\Phi_{c}-\delta\Psi_{c}\right)+\omega_{ti}\left(\delta\Phi_{s}-\delta\Psi_{s}\right)\right]\sin\vartheta_{0}\nonumber \\
 & +i\frac{k_{\vartheta}}{k_{\perp}}\frac{\left(e/m\right)_{i}QF_{0i}}{\omega^{2}-\omega_{ti}^{2}}\left[\left(is\vartheta_{1}\omega-\omega_{ti}\right)\frac{k_{\vartheta}\Omega_{di}}{\omega}\delta\Psi^{\left(0\right)}+i\omega\left(\delta\Phi_{s}-\delta\Psi_{s}\right)-\omega_{ti}\left(\delta\Phi_{c}-\delta\Psi_{c}\right)\right]\cos\vartheta_{0},\label{eq:ki1}
\end{align}
where $N\left(x\right)=\left(1-\frac{\omega_{*ni}}{\omega}\right)\left[x+\left(\frac{1}{2}+x^{2}\right)Z\left(x\right)\right]-\frac{\omega_{*Ti}}{\omega}\left[x\left(\frac{1}{2}+x^{2}\right)+\left(\frac{1}{4}+x^{4}\right)Z\left(x\right)\right]$,
$D\left(x\right)=\left(\frac{1}{x}\right)\left(1+\frac{1}{\tau}\right)+\left(1-\frac{\omega_{*ni}}{\omega}\right)Z\left(x\right)-\frac{\omega_{*Ti}}{\omega}\left[x+\left(x^{2}-\frac{1}{2}\right)Z\left(x\right)\right]$,
$\tau=T_{e}/T_{i}$ and $Z\left(x\right)$ is the plasma dispersion
function. The average of the 1st equation of $\delta K_{E}^{(1)}$
in $\vartheta_{0}$ is already enough for the analysis in this work
and can be given as
\begin{align}
\overline{\omega_{dE}J_{0}\delta K_{E}^{\left(1\right)}} & =\left\langle \frac{e_{E}}{m_{E}}QF_{0E}\frac{k_{\vartheta}}{k_{\perp}}J_{0}^{2}\delta\Psi^{\left(0\right)}\right\rangle +\overline{\frac{e_{E}}{m_{E}}QF_{0}\frac{k_{\vartheta}}{k_{\perp}}\frac{\omega_{dE}}{\omega}J_{0}^{2}\delta\Psi^{\left(1\right)}}+\left\langle QF_{0i}\frac{k_{\vartheta}}{k_{\perp}}\frac{\mu}{k_{\perp}\rho_{E}}2J_{0}J_{1}\delta\hat{B}_{\parallel}^{\left(0\right)}\right\rangle \label{eq:average}
\end{align}
where $\overline{(\cdots)}=\frac{1}{2\pi}\oint\left(\cdots\right)d\vartheta_{0}$.
The average of the 2nd vorticity equation in $\vartheta_{0}$ can
be cast as
\begin{align}
\frac{\partial^{2}}{\partial\vartheta_{1}^{2}}\delta\Psi^{\left(0\right)}+\frac{\omega^{2}}{\omega_{A}^{2}}\left(1-\frac{\omega_{*p}}{\omega}\right)\delta\Psi^{\left(0\right)}+\frac{\omega k_{\vartheta}\alpha R_{0}}{2ck_{\perp}^{2}}\delta\hat{B}_{\parallel}^{\left(0\right)} & =\frac{4\pi\omega}{k_{\perp}k_{\vartheta}c^{2}}q^{2}R_{0}^{2}\overline{\left\langle e_{i}\omega_{di}\delta K_{i}^{\left(1\right)}+e_{E}J_{0}\omega_{dE}\delta K_{E}^{\left(1\right)}\right\rangle }\nonumber \\
+\frac{4\pi}{k_{\perp}^{2}c^{2}}q^{2}R_{0}^{2}\overline{\left\langle \left(1-J_{0}^{2}\right)\omega\omega_{dE}\frac{e_{E}^{2}}{m_{E}}\frac{QF_{0E}}{\omega}\right\rangle \delta\Psi^{\left(1\right)}} & +\frac{4\pi\omega^{2}q^{2}R_{0}^{2}}{k_{\perp}^{2}c^{2}}\left\langle \frac{e_{E}^{2}}{m_{E}}\frac{QF_{0E}}{\omega}\left(1-J_{0}^{2}\right)\right\rangle \delta\Phi^{\left(0\right)}\nonumber \\
 & +\frac{4\pi\omega^{2}q^{2}R_{0}^{2}}{k_{\perp}^{2}c^{2}}\left\langle e_{E}\mu\frac{QF_{0E}}{\omega}\left(1-\frac{2J_{1}J_{0}}{k_{\perp}\rho_{E}}\right)\delta\hat{B}_{\parallel}^{\left(0\right)}\right\rangle .\label{eq:vorticity-2}
\end{align}
By substituting $\delta B_{\parallel}^{\left(0\right)}$, $\delta K_{i}^{\left(1\right)}$,
$\delta K_{E}^{\left(1\right)}$, $\delta\Psi^{\left(1\right)}$ and
Eq.(\ref{eq:average}) into Eq.(\ref{eq:vorticity-2}), the 2nd order
vorticity equation can be re-written as

\begin{equation}
\frac{\partial^{2}}{\partial\vartheta_{1}^{2}}\delta\Psi^{\left(0\right)}+Q\delta\Psi^{\left(0\right)}+\frac{L}{\vartheta_{1}^{2}}\delta\Psi^{\left(0\right)}=0,\label{eq:vorticity-J}
\end{equation}
where $Q\left(\omega\right)=\frac{\omega^{2}}{\omega_{A}^{2}}\left(1-\frac{\omega_{*pi}}{\omega}\right)+q^{2}\frac{\omega\omega_{ti}}{\omega_{A}^{2}}\left[\left(1-\frac{\omega_{*ni}}{\omega}\right)F\left(\frac{\omega}{\omega_{ti}}\right)-\frac{\omega_{*Ti}}{\omega}G\left(\frac{\omega}{\omega_{ti}}\right)-\frac{N^{2}\left(\omega/\omega_{ti}\right)}{D\left(\omega/\omega_{ti}\right)}\right]$,
$F\left(x\right)=x\left(x^{2}+\frac{3}{2}\right)+\left(x^{4}+x^{2}+\frac{1}{2}\right)Z\left(x\right)$,
$G\left(x\right)=x\left(x^{4}+x^{2}+2\right)+\left(x^{6}+\frac{x^{4}}{2}+x^{2}+\frac{3}{4}\right)Z\left(x\right)$,
and
\begin{equation}
L\left(\omega\right)=-\frac{\alpha_{c}\alpha_{E}}{4q^{2}s^{2}}-\frac{\alpha_{E}^{2}}{s^{2}}+\frac{\omega\omega_{*nE}}{s^{2}\omega_{A}^{2}}\frac{n_{E}m_{E}}{k_{\vartheta}^{2}\rho_{tE}^{2}n_{i}m_{i}},
\end{equation}
where $\rho_{tE}^{2}=m_{E}c^{2}T_{E}/e^{2}B^{2}$. As we can see in the epxression of $L$, $L=0$ as $n_E$ goes to zero, which indicates that the vorticity equation goes back to that with only thermal ions\citep{zonca_kinetic_1996}. Note that Eq. (\ref{eq:vorticity-J})
is not periodic anymore and the Floquet theory cannot be applied to
analyze the asymptotic behavior of its solution as before\citep{chen_physics_2016}.
As shown, for $L=0$, Eq.(\ref{eq:vorticity-J}) goes back to the
original vorticity equation in Ref. \citep{zonca_kinetic_1996}. So
the cases that $L\neq0$ are mainly concerned in the rest of this
letter. For $Q=0$, the modes correspond to the accumulation points
in the previous study\citep{zonca_kinetic_1996}. For $Q\text{\ensuremath{\neq0}},$
the solution to Eq.(\ref{eq:vorticity-J}) is 
\begin{equation}
\delta\Psi^{(0)}=\begin{cases}
\sqrt{\left|\vartheta_{1}\right|}I_{d}(h\sqrt{\vartheta_{1}^{2}}), & for\quad\Re h<0\\
\sqrt{\left|\vartheta_{1}\right|}K_{d}(h\sqrt{\vartheta_{1}^{2}}), & for\quad\Re h>0
\end{cases}\label{eq:solutions}
\end{equation}
where $I_{d}$ and $K_{d}$ is d-th order the modified Bessel function
of first and second kind, $h^{2}=-Q$ and $d=\pm\sqrt{1/4-L}$. As
$\left|\vartheta_{1}\right|\rightarrow\infty$, $\sqrt{\left|\vartheta_{1}\right|}I_{d}\sim O\left(e^{\text{\ensuremath{h}}\left|\vartheta_{1}\right|}\right)$
and $\sqrt{\left|\vartheta_{1}\right|}K_{d}\sim O\left(e^{-\text{\ensuremath{h}}\left|\vartheta_{1}\right|}\right)$.
Since the solution should be bounded as $\left|\vartheta_{1}\right|\rightarrow\infty$
and also square integrable to avoid infinite mode energy near resonant
surface, there are two types of physical solutions in Eq. (\ref{eq:solutions}).
The quantity $\Lambda=-i\delta\Psi^{\left(0\right)*}\partial_{\vartheta}\delta\Psi^{(0)}\bigg|{}_{0-}^{0+}$
is of more interest than the explicit solution of $\delta\Psi^{\left(0\right)}$.

\begin{figure}
	\centering
	\includegraphics[totalheight=3in]{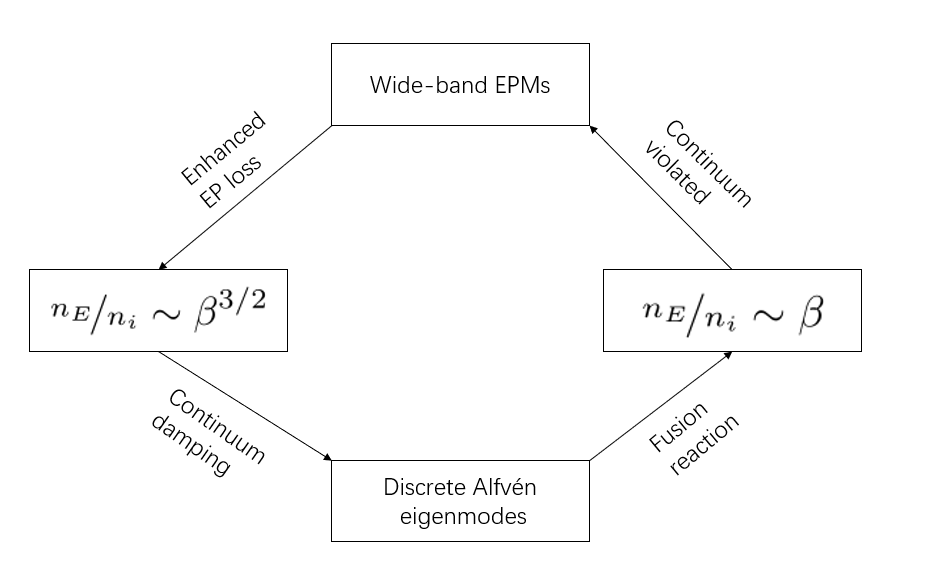}
	\label{fig1}\caption{The illustration of the oscillation mechanism.}		
\end{figure}

For the ideal region,the following quadratic forms are used for matching
the inertial layer and ideal region,

\begin{equation}
\delta W_{k}=\frac{1}{2}\int\delta\Psi_{I}^{*}\frac{4\pi\omega e}{k_{\perp}k_{\vartheta}c^{2}}q^{2}R_{0}^{2}\left\langle \omega_{dE}\delta K_{E,t}\right\rangle d\vartheta\label{eq:Wk}
\end{equation}
\begin{equation}
\delta W_{f}=\frac{1}{2}\int_{-\infty}^{\infty}d\vartheta\left[\left|\frac{d\delta\Psi_{I}}{d\vartheta}\right|^{2}+\left[\frac{k_{\vartheta}^{4}\left(s-\alpha\cos\vartheta\right)^{2}}{k_{\perp}^{4}}-\frac{k_{\vartheta}^{2}\alpha\cos\vartheta}{k_{\perp}^{2}}\right]\left|\delta\Psi_{I}\right|^{2}\right].\label{eq:Wf}
\end{equation}
It should be noted that the EPs contribute mainly to the ideal region due to the small Larmor radius and the fact that the energy of EPs is much higher that that of the thermal ions. By matching the solution of the inertial layer and the ideal region
in integral form, the dispersion relation of concerned modes can be
given as\citep{zonca_kinetic_1996} 

\begin{equation}
i\Lambda=2\delta W_{k}+2\delta W_{f}.\label{eq:dispersion-relation}
\end{equation}
From the equation above, the condition that $\left|\Lambda\right|=\infty$
is impossible to the satisfied. Thus the only physical cases are those
with finite $\Lambda$, which can always be satisfied for $Q\neq0$,
since $Q$ is a complex number in general and either of the two branches
in Eq. (\ref{eq:solutions}) is the physical solution. A spectial
case is that $\Re d=0$. And this alows a band of oscillating modes
with real frequency $\omega<-s^{2}\omega_{A}^{2}k_{\vartheta}^{2}\rho_{E}^{2}n_{i}m_{i}\left(1/4+\alpha_{c}^{2}/4q^{2}s^{2}+\alpha_{E}^{2}/s^{2}\right)/n_{E}m_{E}\omega_{*nE}$.
However, for complex $\omega$, the condition $\Re d\neq0$ can always
be satisfied. Thus, this work is mainly focused on the most unstable
cases that $\Re d\neq0$. If $\Re d>0$, the solution $\sqrt{\left|\vartheta_{1}\right|}I_{d}(h\sqrt{\vartheta_{1}^{2}})$
is stuitable and behaves like $O\left(\left|\vartheta_{1}\right|^{\left|\Re d\right|+1/2}\right)$
as $\left|\vartheta_{1}\right|\rightarrow0$. Otherwise, $\sqrt{\left|\vartheta_{1}\right|}K_{d}(h\sqrt{\vartheta_{1}^{2}})$
is physical solution and also behaves like $O\left(\left|\vartheta_{1}\right|^{\left|\text{\ensuremath{\Re}}d\right|+1/2}\right)$
as $\left|\vartheta_{1}\right|\rightarrow0$. Thus, $\Lambda=0$ as
long as it is finite and $\Re d\neq0$. For marginal stable EPMs,
the dispersion relation for real frequency and the growth rate can
be given as

\begin{equation}
\Re\delta W_{K}\left(\omega_{r}\right)+\delta W_{f}\simeq0,\quad\gamma\simeq-\frac{\Im\delta W_{k}}{\partial\Re\delta W_{k}/\partial\omega_{r}}.\label{eq:disp-marginal}
\end{equation}
In low frequency regime, $\Im\delta W_{k}$ is from the resonance
at precession frequency of trapped EPs. And the real frequency of
mode is determined by non-resonant EPs. Thus, there is no threshold
for EP drive and the equations of non-resonant particles in ideal
region are symmetric for space-time reversing\citep{tsai_theory_1993,chen_theory_1994},
either the $\left(\omega_{r},k_{\vartheta}\right)$ branch or the
$\left(-\omega_{r},-k_{\vartheta}\right)$ branch satisfies the condition
that $-\partial\Re\delta W_{k,u}/\partial\omega_{r}>0$ and becomes
unstable. For instance, a trial function can be assumed as $\delta\Psi_{I}=\kappa_{\perp}/\left(1+s^{2}\vartheta^{2}\right)$,
where $\kappa_{\perp}=k_{\perp}/k_{\vartheta}$. This trial function
and its derivative at $\vartheta\rightarrow\pm\infty$ can match with
$\delta\Psi^{\left(0\right)}$ at $\vartheta\rightarrow0$. A slowing
down distribution of EPs is assumed to be
\begin{equation}
F_{0E}=\frac{3P_{0E}}{4\pi m_{E}\mathcal{E}_{F}}\frac{H\left(\mathcal{E}_{F}-\mathcal{E}\right)}{\left(2\mathcal{E}\right)^{3/2}+\left(2\mathcal{E}_{c}\right)^{3/2}},
\end{equation}
where $H$ is the Heaviside step function, $\mathcal{E}_{c}\ll\mathcal{E}_{F}$
and $\mathcal{E}_{F}$ is the energy of alpha particles in fusion
reaction. The drive part $\Im\delta W_{k,t}$ can be obtained as 
\begin{equation}
\Im\delta W_{k,t}=\Im\frac{1}{2}\int\frac{4\pi\omega e}{k_{\perp}k_{\vartheta}c^{2}}q^{2}R_{0}^{2}\delta\Psi_{I}^{*}\left\langle \omega_{dE}\delta K_{E,t}\right\rangle d\vartheta=\frac{3\pi^{2}\epsilon^{1/2}\alpha_{E}\omega}{16\sqrt{2}\left|s\right|\overline{\omega}_{dF}},
\end{equation}
where $\overline{\omega}_{dF}=-k_{\vartheta}\mathcal{E}_{F}/\Omega_{cE}R$
is the precession frequency evaluated at ${\cal E}_{F}$. Compared
to discrete eigenmodes, the EPMs of this kind are a wider band of
modes and more unstable. From the analysis above, a prediction of
physical phenomenon can be made when the proportion of EPs is near
$O\left(\beta\right)$. As the density of EPs increases, the continuum
structure is regulated and the wide-band EPMs are excited. Thus, a
switch from discrete eigenmodes to wide-band EPMs will appear\citep{Xu_2021}. However,
the collective EPMs are very unstable and the drive is strong. These
modes cause enhanced loss of EPs. If this happens in the igniting
scenario and the EPs production rate is not adequate to overcome the
enhanced loss, the EPs proportion will decrease again. The continuum
structure is then restored and the SAWs again become discrete eigenmodes.
With the fusion reaction and auxiliary heating continuing, the density
of EPs will increase again. Thus, as illustrated in Figure \ref{fig1},
the switch of phases between wide-band EPMs and discrete SAWs can
become a quasi-periodic oscillation, which can possibly be detected
in experiments and simulations. This process may also be affected
by the parametric decay, nonlinear transport and zonal structure\citep{chen_non-linear_2001},
which is beyond the scope of this letter.

In summary, we have presented a linear model for low frequency EPM
in burning tokamak plasmas. It is shown that the continuum structure
is regulated by EPs and a wide-band of EPMs can be excited if the
proportion of EPs is about $O\left(\beta\right)$. Compared with the
analysis before\citep{chen_theory_1994}, the EP drive can excite
the EPMs without overcoming the continuum damping. And the EPMs become
more unstable than discrete eigenmodes. Moreover, a switch or even
an oscillation between discrete eigenmodes and wide-band EPMs can
possibly be detected in future experiments and simulations.

This work currently receives no support of fundings.

\bibliographystyle{vancouver}
\bibliography{MyLibrary}

\begin{thebibliography}{10}

\bibitem{chen_theory_2007}
Chen L, Zonca F.
\newblock Theory of {{Alfv{\'e}n}} waves and energetic particle physics in
  burning plasmas.
\newblock Nuclear Fusion. 2007;47(10):S727-34.

\bibitem{chen_physics_2016}
Chen L, Zonca F.
\newblock Physics of {Alfven} waves and energetic particles in burning plasmas.
\newblock Reviews of Modern Physics. 2016;88(1):015008.

\bibitem{zonca_kinetic_1996}
Zonca F, Chen L, Santoro RA.
\newblock Kinetic theory of low-frequency {{Alfv{\'e}n}} modes in tokamaks.
\newblock Plasma Physics and Controlled Fusion. 1996;38:2011-28.

\bibitem{aymar_iter_2002}
Aymar R, Barabaschi P, Shimomura Y.
\newblock The {ITER} design.
\newblock Plasma Physics and Controlled Fusion. 2002;44(5):519.

\bibitem{heidbrink_observation_1993}
Heidbrink WW, Strait EJ, Chu MS, Turnbull AD.
\newblock Observation of beta-induced {Alfv{\'e}n} eigenmodes in the {DIII}-{D}
  tokamak.
\newblock Phys Rev Lett. 1993;71(6):855-8.

\bibitem{turnbull-global-1993}
Turnbull A, Strait E, Heidbrink W, Chu M, Duong H, Greene J, et~al.
\newblock Global Alfv{\'e}n modes: Theory and experiment.
\newblock Physics of Fluids B: Plasma Physics. 1993;5(7):2546-53.

\bibitem{chen_plasma_1974}
Chen L.
\newblock Plasma heating by spatial resonance of {{Alfv{\'e}n}} wave.
\newblock Physics of Fluids. 1974;17(7):1399.

\bibitem{chen_theory_1994}
Chen L.
\newblock Theory of magnetohydrodynamic instabilities excited by energetic
  particles in tokamaks.
\newblock Physics of Plasmas. 1994 May;1(5):1519-22.

\bibitem{Tang_kinetic_1980}
Tang WM, Connor JW, Hastie RJ.
\newblock Kinetic-ballooning-mode theory in general geometry.
\newblock Nuclear Fusion. 1980;20(11):1439.

\bibitem{cheng_kinetic_1982}
Cheng CZ.
\newblock Kinetic theory of collisionless ballooning modes.
\newblock Physics of Fluids. 1982;25(6):1020.

\bibitem{chen_kinetic_1991}
Chen L, Hasegawa A.
\newblock Kinetic {Theory} of {Geomagnetic} {Pulsations} {Internal}
  {Excitations} by {Energetic} {Particles}.
\newblock Journal of Geophysical Research. 1991;96(A2):1503-12.

\bibitem{antonsen_kinetic_1980}
Antonsen TM, Lane B.
\newblock Kinetic equations for low frequency instabilities in inhomogeneous
  plasmas.
\newblock Physics of Fluids. 1980;23(6):1205.

\bibitem{connor_shear_1978}
Connor JW, Hastie RJ, Taylor JB.
\newblock Shear, {Periodicity}, and {Plasma} {Ballooning} {Modes}.
\newblock Phys Rev Lett. 1978;40(6):394.

\bibitem{tsai_theory_1993}
Tsai ST, Chen L.
\newblock Theory of kinetic ballooning modes excited by energetic particles in
  tokamaks.
\newblock Physics of Fluids B: Plasma Physics. 1993;5(9):3284.

\bibitem{Xu_2021}
Xu L, Shen W, Ren Z, Zhou T, Wang Y, Hui L, et~al.
\newblock Investigation of Alfv{\'e}n eigenmodes and energetic particle modes
  in EAST with neutral beam injection.
\newblock Nuclear Fusion. 2021;61(7):076005.

\bibitem{chen_non-linear_2001}
Chen L, Lin Z, White RB, Zonca F.
\newblock Non-linear zonal dynamics of drift and drift-{{Alfv{\'e}n}}
  turbulence in tokamak plasmas.
\newblock Nuclear Fusion. 2001;41(6).

\end{thebibliography}

\end{document}